\documentclass[twoside,12pt]{article}
\usepackage{CJK}
\usepackage{indentfirst}
\usepackage{bm}
\usepackage{epsfig}
\usepackage{amsmath}
\usepackage{graphicx}
\usepackage{multirow}
\usepackage{dcolumn}
\usepackage{booktabs}
\usepackage{xcolor}
\usepackage{cite}
\usepackage{hyperref}
\usepackage{comment}
\usepackage{physics}

\begin{document}


\thispagestyle{empty} \vspace*{0.8cm}\hbox
to\textwidth{\vbox{\hfill\huge\sf Commun. Theor. Phys.\hfill}}
\par\noindent\rule[3mm]{\textwidth}{0.2pt}\hspace*{-\textwidth}\noindent
\rule[2.5mm]{\textwidth}{0.2pt}


\begin{center}
\LARGE\bf Magnetization-resolved density of states and quasi-first order transition in the two-dimensional random bond Ising model: an entropic sampling study
\end{center}

\footnotetext{\hspace*{-.45cm}\footnotesize $^*$E-mail: yaodaox@mail.sysu.edu.cn}
\footnotetext{\hspace*{-.45cm}\footnotesize $^\dag$E-mail: tangleihan@westlake.edu.cn}

\begin{center}
\rm Yi Liu$^{\rm a)}$, \ \ Ding Wang$^{\rm b,c)}$, \ \ Xin Wang$^{\rm a)}$, \ \ Dao-Xin Yao$^{\rm a, ^*) }$, \ and  \ Lei-Han Tang$^{\rm b, ^\dag)}$
\end{center}

\begin{center}
\begin{footnotesize} \sl
Center for Neutron Science and Technology, Guangdong Provincial Key Laboratory of Magnetoelectric Physics and Devices, School of Physics, Sun Yat-Sen University, Guangzhou 510275, China$^{\rm a)}$ \\

Center for Interdisciplinary Studies and Department of Physics, Westlake University, Hangzhou 310030, China$^{\rm b)}$ \\

Department of Physics, Hong Kong Baptist University, Kowloon Tong, Hong Kong SAR, China$^{\rm c)}$ \\
\end{footnotesize}
\end{center}

\begin{center}

\end{center}

\vspace*{2mm}

\begin{center}
\begin{minipage}{15.5cm}

\parindent 20pt\footnotesize
Systems with quenched disorder possess complex energy landscapes that are challenging to explore under conventional Monte Carlo method. 
In this work, we implement an efficient entropy sampling scheme for accurate
computation of the entropy function in low-energy regions. The method is applied to the two-dimensional $\pm J$ random-bond Ising model, where frustration is controlled by the fraction $p$ of ferromagnetic bonds. We investigate the low-temperature paramagnetic--ferromagnetic phase boundary below the multicritical point at $T_N = 0.9530(4)$, $P_N = 0.89078(8)$, as well as the zero-temperature ferromagnetic--spin-glass transition. Finite-size scaling analysis reveals that the phase boundary for $T < T_N$ exhibits reentrant behavior. By analyzing the evolution of the magnetization-resolved density of states $g(E, M)$ and ground-state spin configurations against increasing frustration, we provide strong evidence that the zero-temperature transition is quasi–first order. 
Finite-size scaling conducted on the spin-glass side supports the validity of $\beta = 0$, with a correlation length exponent $\nu = 1.50(8)$. 
Our results provide new insights into the nature of the ferromagnetic-to-spin-glass phase transition in an extensively degenerate ground state.
\end{minipage}
\end{center}

\begin{center}
\begin{minipage}{15.5cm}
\begin{minipage}[t]{2.3cm}{\bf Keywords:}\end{minipage}
\begin{minipage}[t]{13.1cm}
Random bond Ising model,  entropic sampling, quasi-first order transition
\end{minipage}\par\vglue8pt

\end{minipage}
\end{center}

\section{Introduction}
\begin{figure}[t]
    \centering
    \includegraphics[width=0.5\linewidth]{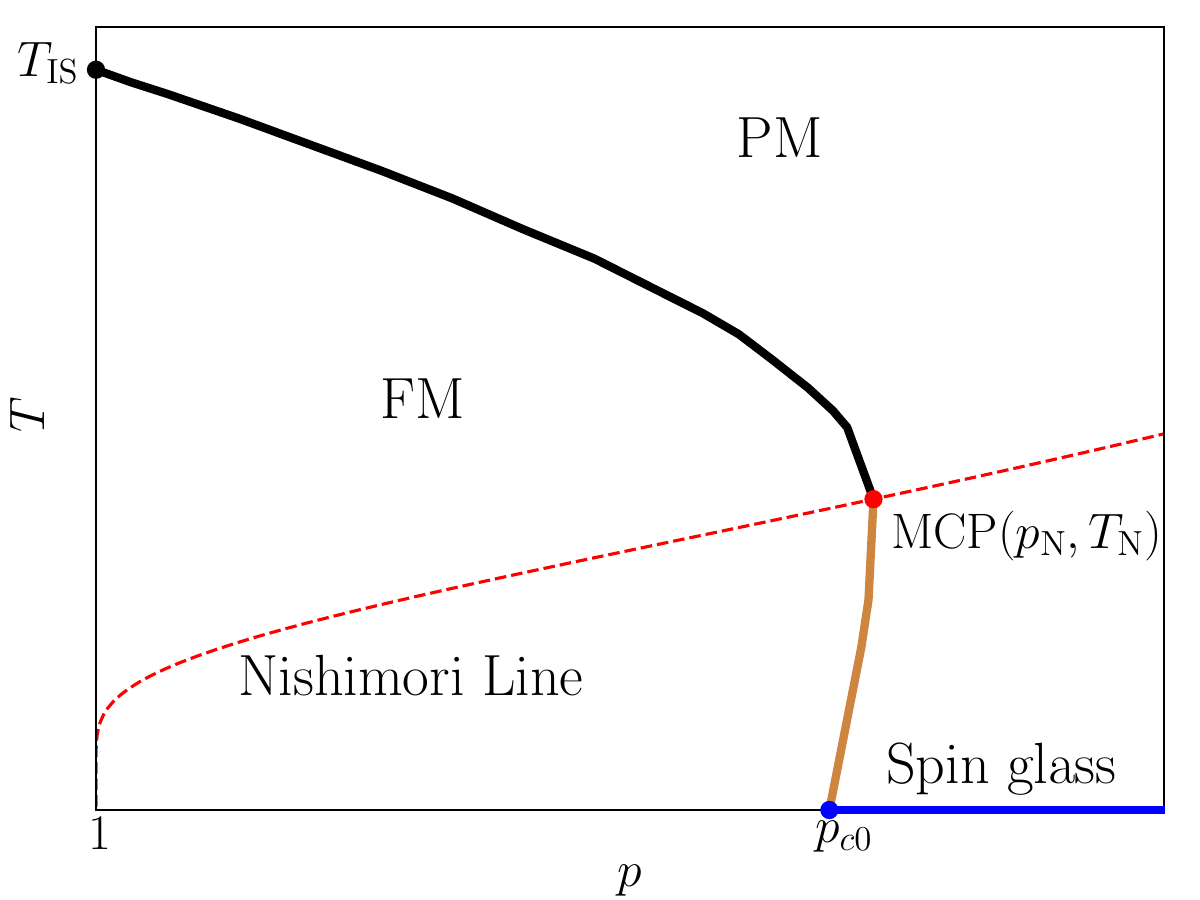}
    \caption{Phase diagram of the 2D $\pm J$ RBIM. The ferromagnetic phase boundary at high temperatures (black line) starts from the pure case at $T_{\text {IS}}/J=2.269...$ and ends at the multicritical point on the Nishimori line (red dashed line). Below the multicritical point, the phase boundary curves slightly to the left and reaches the zero-temperature ferromagnetic to spin-glass transition point at $p_{c0}\simeq 0.9$. Thus between $p_{\rm N}$ and $p_{c0}$, the paramagnetic phase ``re-enters'' at sufficiently low temperatures.
    These phase boundaries were drawn based on data from Refs.~\cite{merz2002two,amoruso2004domain,parisen2009strong}.
    }
    \label{fig:phase diagram}
\end{figure}

The study of disordered magnetic systems has attracted much attention in statistical physics due to the rich physics associated with frustration that
disrupts an otherwise ordered state. Among various models, the $\pm J$ random‐bond Ising model (RBIM) stands out as a classic example. Its Hamiltonian is given by
\begin{equation}
    H(\{S_i\})=-\sum_{\left \langle ij \right \rangle}J_{ij}S_iS_j,
\end{equation}
where $S_i=\pm 1, i=1,\ldots, N$ are the Ising spins associated with the lattice sites, $\left \langle ij \right \rangle$ denotes all nearest neighbor pairs, and $J_{ij}$ are quenched random variables chosen independently from a bimodal distribution:
\begin{equation}
    P(J_{ij})=p\delta(J_{ij}-J)+(1-p)\delta(J_{ij}+J).
\end{equation}
In this work, we shall focus exclusively on the two-dimensional (2D) square lattice with $J=1$ and periodic boundary conditions. 


Extensive work has been done to map the phase diagram of the 2D $\pm J$ RBIM on the $p-T$ plane as shown in figure~\ref{fig:phase diagram} (see, e.g., Refs.~\cite{merz2002two,parisen2009strong,amoruso2004domain,thomas2011simplest}). 
There exists a special line, known as the Nishimori line (NL)~\cite{nishimori1980exact, nishimori1981internal, nishimori2001statistical}, along which the energy and specific heat as a function of temperature $T$ and ferromagnetic (FM) bond percentage $p$ can be computed exactly. The NL is given by
\begin{equation}
    e^{2J/T}=\frac{p}{1-p}.
\end{equation}
It crosses the ferromagnetic phase boundary at the multicritical point (MCP), separating it into upper (Ising‐like) and lower parts. 
While transition on the high temperature side is governed by the 2D Ising fixed point, with disorder playing only a minor role~\cite{hasenbusch2008universal, shalaev1994critical, shankar1987exact}, its nature on the low-temperature side remains controversial~\cite{amoruso2004domain, McMillan1984domain, kirkpatrick1977frustration, vannimenus1977theory}. 

Conventional Monte Carlo methods are inefficient in equilibrating frustrated systems at low temperatures \cite{Landau2014Guide,binder1986spin}. Enhanced sampling algorithms, such as replica exchange Monte Carlo method~\cite{hukushimaexchange,swendsen_replica_exchanges,Marinari_1992}, show better performance, but still have difficulty accessing true ground states. In contrast, exact matching algorithms can identify individual ground‑state configurations of the 2D $\pm J$ RBIM on a planar graph \cite{amoruso2004domain}. Finite-size scaling (FSS) of the resulting ground‑state properties yields a zero‑temperature critical point at $p_{c0} \simeq 0.897$. However, since the degeneracy of the ground-state grows exponentially with the size of the system, a more complete sampling of ground state configurations is required to determine relevant statistics, especially around the FM phase boundary at zero and low temperatures.

Entropic sampling methods~\cite{berg1991multicanonical, berg1992multicanonical, berg1992new, Lee1993New}, which iteratively build the density of states (DOS) of the system by biasing Monte Carlo moves to flatten the energy histogram, offer a direct route to ground‐state entropy. However, the original Berg's scheme~\cite{berg1992new}, which has been successful in treating first-order phase transitions, converges slowly towards the ground state. 
Wang and Landau~\cite{wang2001determining, wang2001efficient} (WL) introduced a continuous iterative scheme that gained wide acceptance. 
An alternative algorithm to accelerate DOS convergence through extrapolation was implemented by Tang~\cite{Tang2006}. The latter has been shown to be effective in treating frustrated systems~\cite{Tang2004Dir, tang2005zero} and is adopted in this study. Enhanced with noise filtering and DOS extrapolation, this method delivers high‑precision estimates of the ground‑state entropy across a wide range of $p$ values. By careful examination of how ground-state DOS changes as $p$ decreases, we found compelling evidence for a quasi-first order transition from the FM to the spin-glass (SG) phase. Our results also shed light on the shape of the ferromagnetic phase boundary at low temperatures.

The paper is organized as follows. 
In section~\ref{sec:algorithm}, we present the details of the entropy sampling algorithm used in our work.
In section~\ref{sec:fss_results}, we apply the FSS analysis to the Binder cumulant and the disorder-averaged magnetization to map out the phase boundary below MCP. 
Section~\ref{sec:zero_temperature} presents the analyses of zero-temperature entropy and magnetization, revealing first-order–like behavior.
Conclusions are drawn in section~\ref{sec:discussion}.
Appendix~\ref{sec:simulation_detail} provides the simulation details.

\section{Entropic sampling}
\label{sec:algorithm}

The Metropolis Monte Carlo algorithm samples spin configurations $\{S_i\}$ according to the Boltzmann weight $P(E)\sim e^{-E/T}$, where $E$ is the total energy of the system. Each proposed spin flip that changes the system energy from $E_1$ to $E_2$ is accepted with the probability
\begin{equation}
    P(E_1 \to E_2) = \min\left(\frac{P(E_2)}{P(E_1)}, 1\right).
    \label{eq:importance}
\end{equation}
In entropic sampling, one adopts instead 
\begin{equation}
    P(E)\sim 1/g(E)=e^{-S(E)}
    \label{eq:entropic-sampling-weight}
\end{equation} 
where $g(E)$ is the density of states at energy $E$ and $S(E)=\ln g(E)$ the system entropy. The spin flips are performed again following equation~(\ref{eq:importance}). This procedure yields a flat energy histogram when running the simulation for a long enough time. 

As $g(E)$ is not known {\it a priori}, Berg introduced an iterative procedure that starts with a trial sampling weight $P_0(E)$. In each round of simulation, the energy histogram $h_k(E)$ is collected over a sufficiently long period, from which  
the sampling weight for the next round is constructed according to
\begin{equation}
    P_{k+1}(E)={P_{k}(E)\over h_k(E)+a},\qquad\qquad k=0,1,\ldots
    \label{eq:update1}
\end{equation}
where a ``floor value'' $a\sim 1$ is introduced for $E$ values with insufficient statistics. The above equation can be rewritten in logarithmic form,
\begin{align}
    R_{k+1}(E)=R_{k}(E) + \ln \bigl[h_k(E)+a\bigr],
    \label{eq:update2}
\end{align}
where $R_k(E)=-\ln{P_{k}(E)}$ is the estimated entropy function up to a constant.

\begin{figure}[t]
    \centering
    \includegraphics[width=0.49\linewidth]{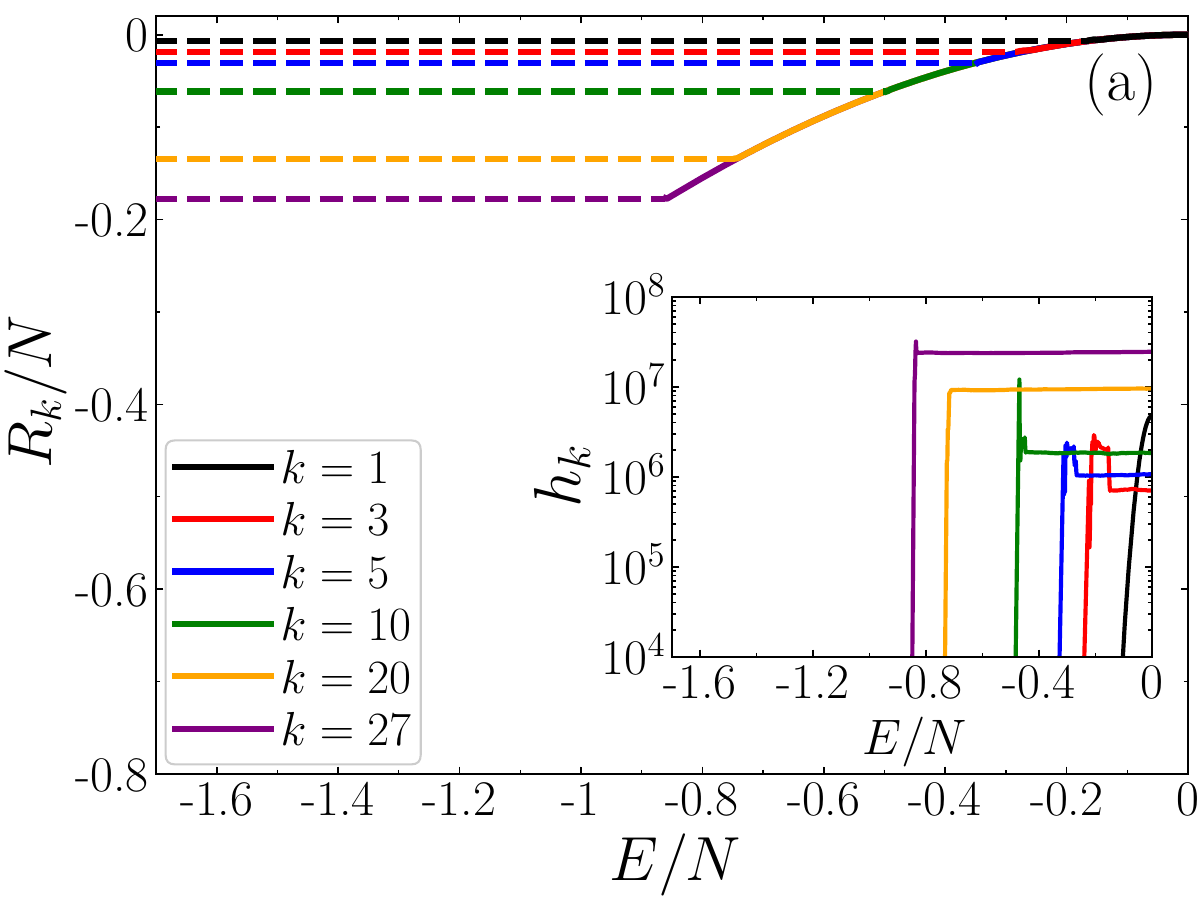} 
    \includegraphics[width=0.49\linewidth]{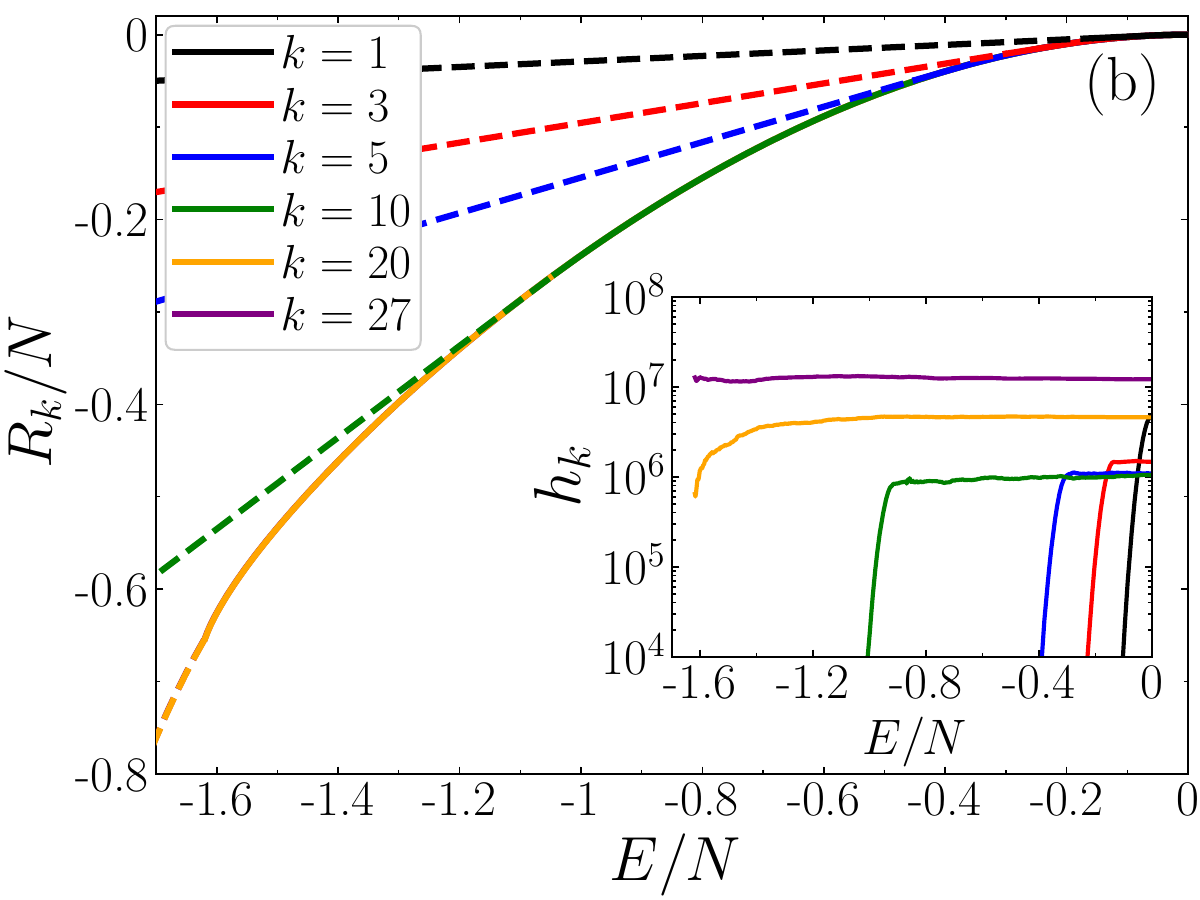} 
    \caption{Comparison of two iteration schemes for the 2D $\pm J$ RBIM with $N=48 \times 48$ and $p=0.9$. (a) Berg’s iterative scheme, (b) the polynomial fitting and extrapolation scheme. Both panels show the evolution of relative entropy and histogram as a function of the number of iterations. The dashed lines indicate regions where \(h_k = 0\), while the solid lines correspond to regions where \(h_k > 0\). The polynomial fitting and extrapolation scheme exhibits significantly faster convergence compared to Berg’s original method.}
    \label{fig:diff it schemes}
\end{figure}

Figure~\ref{fig:diff it schemes}(a) shows $R_k(E)$ obtained from simulations of the 2D $\pm J$ RBIM on a $48\times 48$ square lattice at $p=0.9$, starting from $P_0(E)= const.$ that corresponds to the Boltzmann distribution at $T=\infty$. The total number of trial spin flips increases by a factor of 1.2 in successive iterations, with measured histograms $h_k(E)$ are shown in the insert. Upon sufficient sampling, we have
\begin{equation}
h_k(E)\sim g(E)P_{k}(E)=e^{S(E)-R_{k}(E)}.
\label{eq:h_k-measured}
\end{equation}
Let $E_{k,\rm min}$ be the value of $E$ where $R_{k}(E)$ switches from $S(E)$ to a constant value, as shown in figure ~\ref{fig:diff it schemes}(a). The condition $h_k(E_{k, \rm min}-\Delta E_{k,\rm min})=a$ yields
\begin{equation}
\Delta E_{k,\rm min}\equiv E_{k,\rm min}-E_{k+1,\rm min}\simeq T_k\ln (\bar{h}/a),
\label{eq:Berg-DeltaE}
\end{equation}
where $T_k^{-1}=\partial S(E_{k,\rm min})/\partial E$ and $\bar{h}=h_k(E_{k,\rm min})$.
Thus, the iterative scheme (\ref{eq:update2}) becomes increasingly inefficient as one enters the low temperature regime.

To overcome this difficulty, we perform a least-squares fit of the right-hand-side of equation~\eqref{eq:update2} to a polynomial over a range of $E$ values around 
$E_{k,\rm min}$ where sufficient statistics is warranted. This polynomial function is used to construct a trial $R_{k+1}(E)$ up to a new $E_{k+1, \rm min}<E_{k,\rm min}$, beyond which $R_{k+1}(E)$ is smoothly connected to a linear function of $E$ at temperature $T_{k+1}<T_k$. This procedure produces results shown in figure~\ref{fig:diff it schemes}(b), with the length of simulation in each round identical to the previous case. It is evident that this modified scheme is much more efficient in reaching the low energy states of the model.


It should be noted that, as one approaches the ground state under a given bond configuration, the entropy function $S(E)$ may develop structures not well approximated by a polynomial of low degree. Therefore, once we have reached energies quite close to the ground state, we switch back to equation~\eqref{eq:update2} for further iteration and convergence. Furthermore, to examine in greater detail the degenerate ground states with different magnetization $M=\sum_i S_i$, we extended the flat-histogram scheme to the 2D $(E,M)$ plane~\cite{wang2001determining, Landau2004approach}), building the joint DOS \( g(E, M) \) using a 2D form of equation~\eqref{eq:update2}. 

The above procedure allows us to sample the entire energy range for systems up to $64\times 64$. Since equation~\eqref{eq:h_k-measured} only yields the entropy function $S(E)=\ln g(E)$ up to a constant, we use the condition
\begin{equation}
\sum_E g(E)=2^N
\end{equation}
to compute the absolute value of $g(E)$ and $S(E)$. The 2D density of states $g(E,M)$ satisfies
$\sum_M g(E,M)=g(E)$. We have also imposed the symmetry $h_k(E,-M)=h_k(E,M)$ in constructing the 2D histogram.

\section{Ferromagnetic phase boundary and critical exponents}
\label{sec:fss_results}

In this section, we examine the critical properties of magnetization along the phase boundary below the MCP. We apply the FSS ansatz to the high-precision data generated under the entropic sampling scheme.
Despite the relatively small range of system sizes, accurate estimates of the transition line and the critical exponents are obtained,
in excellent agreement with previous studies using a variety of numerical methods.

Table~\ref{tab:summary_below_MCP} shows a set of representative results over the years. Most studies yield a critical value $p_{c0}$ between $0.895$ to $0.897$ for the zero temperature transition, but a slightly lower yet different value of $p_c$ at $T=0.5$. The critical value at MCP is estimated to be  $p_{\rm N}=0.8908$~\cite{hasenbusch2008multicritical, parisen2009strong, Chen2025Tensor}.

\begin{table}[t]
    \centering
    \footnotesize
    \begin{tabular}{l c c c c c c}
        \toprule
        \textbf{Method} & \textbf{Year} & \textbf{Temperature} & $L_{\max}$ & $p_c$ & $\nu$ & $\beta$ \\
        \midrule
        Series expansion~\cite{grinstein1979ising} & 1979 & 0 & -- & $\sim 0.901$ & -- & -- \\
        Matching algorithm~\cite{Freund_1989} & 1989 & 0 & 210 & 0.895(10) & -- & -- \\
        Matching algorithm~\cite{BENDISCH1994139} & 1994 & 0 & 300 & 0.892$\sim$0.905 & -- & -- \\
        Exact Ground States~\cite{Kawashima_1997} & 1997& 0 & 32 & 0.896(1) & 1.30(2) & -- \\
        \multirow{2}{*}{Real-space renormalization group~\cite{nobre2001phase}} & \multirow{2}{*}{2001} & 0 & -- & 0.8951(3) & -- & -- \\
        & & 0.5 & -- & 0.8919(4) & -- & -- \\
        Exact Ground States~\cite{wang2003confinement} & 2003 & 0 & 42 & 0.8967(1) & 1.49(2) & -- \\
        Matching algorithm~\cite{amoruso2004domain} & 2004 & 0 & 700 & 0.897(1) & 1.55(1) & 0.09(1) \\
        \multirow{2}{*}{Monte Carlo~\cite{parisen2009strong}} & \multirow{2}{*}{2009} & 0.5 & 64 & 0.8925(1) & 1.50(4) & 0.092(2) \\
        & & 0.645 & 64 & 0.8915(2) & 1.50(4) & 0.099(3) \\
        Pfaffian technique~\cite{thomas2011simplest} & 2011 & below MCP & 512 & -- & 1.52(5) & -- \\
        \bottomrule
    \end{tabular}
    \caption{Summary of critical point and critical exponent estimates below the MCP for the 2D $\pm J$ RBIM on the square lattice. In the table, $L_{\max}$ denotes the largest system size examined in each study, $p_c$ is the estimated critical point, $\nu$ is the correlation length exponent, and $\beta$ is the magnetization exponent.}
    \label{tab:summary_below_MCP}
\end{table}

In entropic sampling, a single simulation is sufficient to obtain the temperature dependence of any physical quantity $X$ for a given bond configuration. One simply keeps track of the mean value $X(E)$ of $X$ over configurations at energy $E$, and use the reweighting formular
\begin{equation}
    \expval{X}(T)=\frac{\sum_{E}X(E)g(E)e^{-E/T}}{\sum_E g(E)e^{-E/T}},
\end{equation}
at the end of the simulation. The result is further averaged over different disorder realizations, denoted as \( [\expval{X}](T) \), where \( [\cdots] \) denotes the disorder average. 

In the case of total magnetization $M$, an alternative way is to first collect the 2D histogram $g(E,M)$ during simulation. Its moments can then be computed according to
\begin{equation}
    \expval{M^n}(T)=\frac{\sum_{E,M} |M^n| g(E,M)e^{-E/T}}{\sum_{E,M} g(E,M)e^{-E/T}},
\end{equation}
followed by average over disorder realizations. The Binder cumulant
\begin{equation}
    B(T)=\frac{[\expval{M^4}]}{[\expval{M^2}]^2}
\end{equation}
can be calculated this way. Additional details of our implementation can be found in Appendix~\ref{sec:simulation_detail}.

Figure~\ref{fig:Binder at diff T} shows \( B(T) \) at four selected temperatures $T$. In all cases, data at different $L$ intersect at a critical value $p_c(T)$. Apart from the value $0.8945$ at $T=0$, which is somewhat lower, the rest are in good agreement with the values listed in Table 1.

\begin{figure*}[t]
    \centering
    \includegraphics[width=\linewidth]{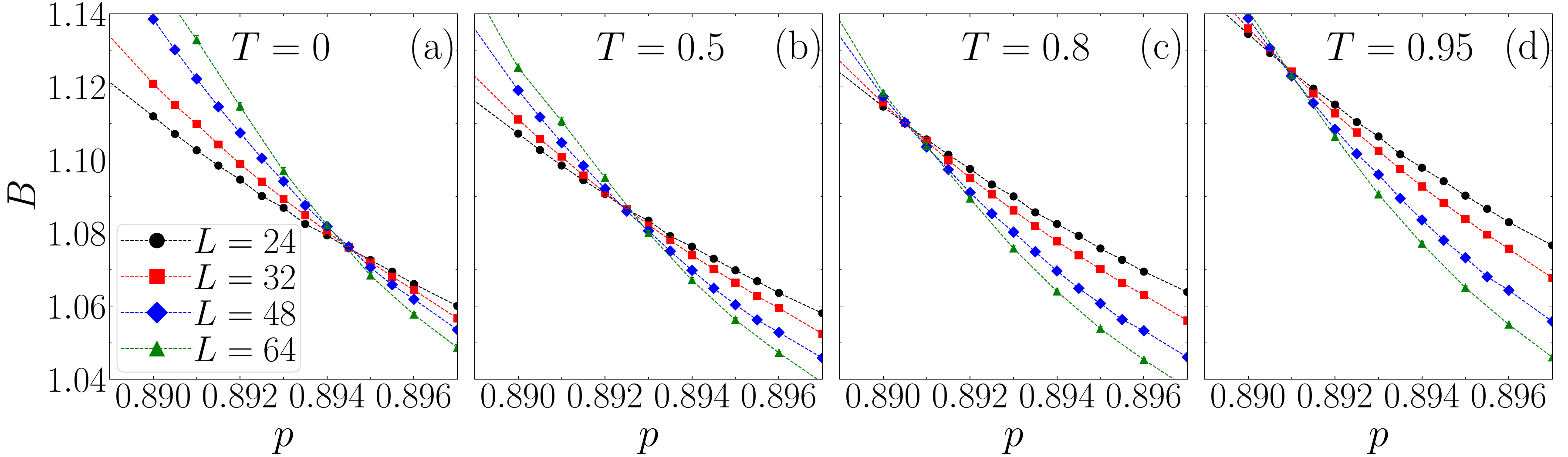}
    \caption{Binder cumulant $B$ as a function of $p$ below the MCP. Panels (a)–(d) correspond to temperatures $T = 0$, $0.5$, $0.8$, and $0.95$, respectively.}
    \label{fig:Binder at diff T}
\end{figure*}

A more systematic approach to estimate $p_c(T)$ and the finite-size scaling exponent $\nu(T)$ is to fit the Binder cumulant data at a given $T$, but different sizes $L$ and FM bond concentration $p$ to the scaling form,
\begin{equation}
    B(L,p) = \tilde{B} \left( (p-p_c)L^{1/\nu} \right).
    \label{eq:B_scaling}
\end{equation}
In Refs.~\cite{hasenbusch2008multicritical, parisen2009strong, Chen2025Tensor},
the quality of the fit is assessed using the reduced chi-square static, \( \chi^2/\text{DOF} \), where a good fit corresponds to \( \chi^2/\text{DOF} \approx 1 \). Here, DOF stands for the number of degrees of freedom in the fit. 
In our implementation, the scaling function $\tilde{B}(x)$ is approximated by a polynomial of degree $n_{\rm max}=3$,
\begin{equation}
    \tilde{B} = B_0 + \sum_{n=1}^{n_{\max}} B_n x^n,
    \label{eq:B_fitting}
\end{equation}
and we perform a least-squares fit to the Binder cumulant data , treating 
\( B_0 \), \( \{B_n\} \), \( p_c \), and \( \nu \) as fitting parameters. 

Including data from very small system sizes degrades the quality of the fit, often resulting in \( \chi^2/\text{DOF} \gg 1 \). As the minimum system size $L_\text{min}$ increases, $\chi^2/\text{DOF}$ decreases and eventually approaches 1. We select \( L_\text{min} \) as the smallest system size beyond which further increases do not significantly reduce \( \chi^2/\text{DOF} \). 
To estimate statistical uncertainties of the fitted parameters, we employ the bootstrap method. Throughout this study, we report uncertainties at the $3\sigma$ (99.7\%) confidence level, unless otherwise stated.

Following the above procedure, we obtain estimates of the critical parameters as summarized in Table~\ref{tab:B_scaling}. The correlation length exponent \( \nu \) below the MCP is about 1.5, independent of $T$. This is in agreement with previous findings~\cite{wang2003confinement, amoruso2004domain, parisen2009strong, thomas2011simplest}. 


\begin{table}[t]
\centering
\caption{Fitting results for the FSS of the Binder cumulant \( B \) at different temperature using equation~\eqref{eq:B_fitting}. $T_{\text{NL}}$ represents the temperature along the NL, indicating that the $B$ data pass through the MCP along the NL.}
\label{tab:B_scaling}
\begin{tabular}{ccccc}
Temperature & $\chi^2/\text{DOF}$ & $p_c$ & $\nu$ & $B_0$ \\  
 \hline
$0$    &  $47.7 / 40$  &  $0.89456(13)$  &  $1.50(8)$  &  $1.0755(10)$  \\
$0.1$  &  $47.7 / 40$  &  $0.89456(13)$  &  $1.50(8)$  &  $1.0755(10)$  \\
$0.2$  &  $47.8 / 40$  &  $0.89456(13)$  &  $1.50(8)$  &  $1.0755(10)$  \\
$0.3$  &  $47.7 / 40$  &  $0.89426(12)$  &  $1.50(8)$  &  $1.0774(9)$  \\
$0.4$  &  $29.3 / 40$  &  $0.89327(7)$   &  $1.54(7)$  &  $1.0831(8)$  \\
$0.5$  &  $33.5 / 40$  &  $0.89233(9)$   &  $1.55(8)$  &  $1.0882(9)$  \\
$0.6$  &  $36.0 / 40$  &  $0.89155(10)$  &  $1.55(7)$  &  $1.0937(9)$  \\
$0.65$ &  $36.4 / 40$  &  $0.89122(10)$  &  $1.56(6)$  &  $1.0969(9)$  \\
$0.7$  &  $36.4 / 40$  &  $0.89094(10)$  &  $1.56(6)$  &  $1.1006(10)$  \\
$0.8$  &  $35.7 / 40$  &  $0.89055(10)$  &  $1.55(6)$  &  $1.1096(10)$  \\
$0.9$  &  $35.2 / 40$  &  $0.89054(9)$   &  $1.53(5)$  &  $1.1207(10)$  \\
$0.95$ &  $36.2 / 40$  &  $0.89075(8)$   &  $1.51(5)$  &  $1.1269(9)$  \\
$T_{\text{NL}}$  &  $36.3 / 40$  &  $0.89078(8)$  &  $1.58(5)$  &  $1.1273(10)$  \\
\end{tabular}
\end{table}

With \( p_c \) and \( \nu \) determined above, we analyze the mean magnetization $m=\langle M\rangle/N$ across the phase boundary.
Assuming the FSS ansatz, 
\begin{equation}
    m(L,p) = L^{-\beta / \nu} \bar{u}_h(p) \tilde{m} \left( (p-p_c)L^{1/\nu} \right),
    \label{eq:M_scaling}
\end{equation}
where \( \beta  \) (not to be confused with the inverse temperature) is the magnetization exponent at a given $T$, $\bar{u}_h(p)$ is a function of $p$ related to the magnetic nonlinear scaling field~\cite{parisen2009strong}, and $\tilde{m}$ is the scaling function. According to this FSS form, we fit the numerical data using
\begin{equation}
    \ln m = -\frac{\beta}{\nu} \ln L 
    + \sum_{n=0}^{n_{\max}} a_n (p - p_c)^n L^{n/\nu} 
    + \sum_{m=0}^{m_{\max}} b_m (p - p_c)^m.
    \label{eq:M_fitting}
\end{equation}
The last term represents the analytic corrections \cite{parisen2009strong}, and its inclusion significantly reduces \( \chi^2/\text{DOF} \), bringing it closer to 1. In the fitting, we set \( n_{\max} = 3 \) and \( m_{\max} = 1 \), and fix \( p_c \) using the value listed in Table~\ref{tab:B_scaling}. The fitting results are summarized in Table~\ref{tab:M_scaling}. Our estimates for \( \beta \) are in close agreement with the results reported in Ref.~\cite{parisen2009strong}: \( \beta = 0.092(2) \) at \( T = 0.5 \), and \( \beta = 0.099(3) \) at \( T = 0.645 \); as well as with those in Ref.~\cite{amoruso2004domain}: \( \beta = 0.09(1) \) at \( T = 0 \).

\begin{table}[t]
\centering
\caption{Fitting results for the FSS of the magnetization \( m \) at different temperature, using the equation~\eqref{eq:M_fitting}. $T_{\text{NL}}$ represents the temperature along the NL, indicating that the $m$ data pass through the MCP along the NL.}
\label{tab:M_scaling}
\begin{tabular}{cccc}
Temperature & $\chi^2/\text{DOF}$ & $\beta / \nu$ & $\beta$  \\  
 \hline
$0$    &  $45.0 / 40$  &  $0.0552(7)$  &  $0.083(5)$  \\
$0.1$  &  $45.0 / 40$  &  $0.0552(7)$  &  $0.083(5)$  \\
$0.2$  &  $45.0 / 40$  &  $0.0552(7)$  &  $0.083(5)$  \\
$0.3$  &  $43.8 / 40$  &  $0.0563(6)$  &  $0.085(5)$  \\
$0.4$  &  $34.4 / 40$  &  $0.0598(5)$  &  $0.091(5)$  \\
$0.5$  &  $40.4 / 40$  &  $0.0631(7)$  &  $0.097(5)$  \\
$0.6$  &  $44.7 / 40$  &  $0.0666(8)$  &  $0.103(5)$  \\
$0.65$ &  $46.4 / 40$  &  $0.0688(8)$  &  $0.107(5)$  \\
$0.7$  &  $47.8 / 40$  &  $0.0712(8)$  &  $0.109(5)$  \\
$0.8$  &  $49.4 / 40$  &  $0.0771(9)$  &  $0.120(5)$  \\
$0.9$  &  $48.8 / 40$  &  $0.0843(9)$  &  $0.130(5)$  \\
$0.95$ &  $48.0 / 40$  &  $0.0883(8)$  &  $0.134(5)$  \\
$T_{\text{NL}}$  &  $50.8 / 40$  &  $0.0886(9)$  &  $0.141(5)$  \\
\end{tabular}
\end{table}


\section{Ground state and quasi-first order transition}
\label{sec:zero_temperature}

In section~\ref{sec:fss_results}, we analyzed the behavior of system magnetization near the FM phase boundary within the framework of FSS, under the implicit assumption that the transition is continuous. However, the small value of $\beta/\nu$ obtained could also be consistent with a first-order-like transition, where the total magnetization exhibits a discontinuous jump rather than vanishing continuously as the fraction of antiferromagnetic (AFM) bonds increases. In this section, we explore evidence supporting this alternative scenario.



\subsection{Magnetization-resolved degeneracy of low-lying states}
\label{sec:reentrant_sample}


We begin by examining how the magnetization profile of the ground state and low-lying excited states evolves as AFM bonds are gradually introduced into the system. Let $N_a$ denote the number of AFM bonds, which is initially set to 0. At each step, a FM bond is randomly selected and converted into an AFM one, thereby increasing $N_a$ by one. We then compute the joint DOS \( g(E, M) \) of the system, from which the mean magnetization at a given $E$ is obtained.

\begin{figure*}[t]
    \centering
    \includegraphics[width=\linewidth]{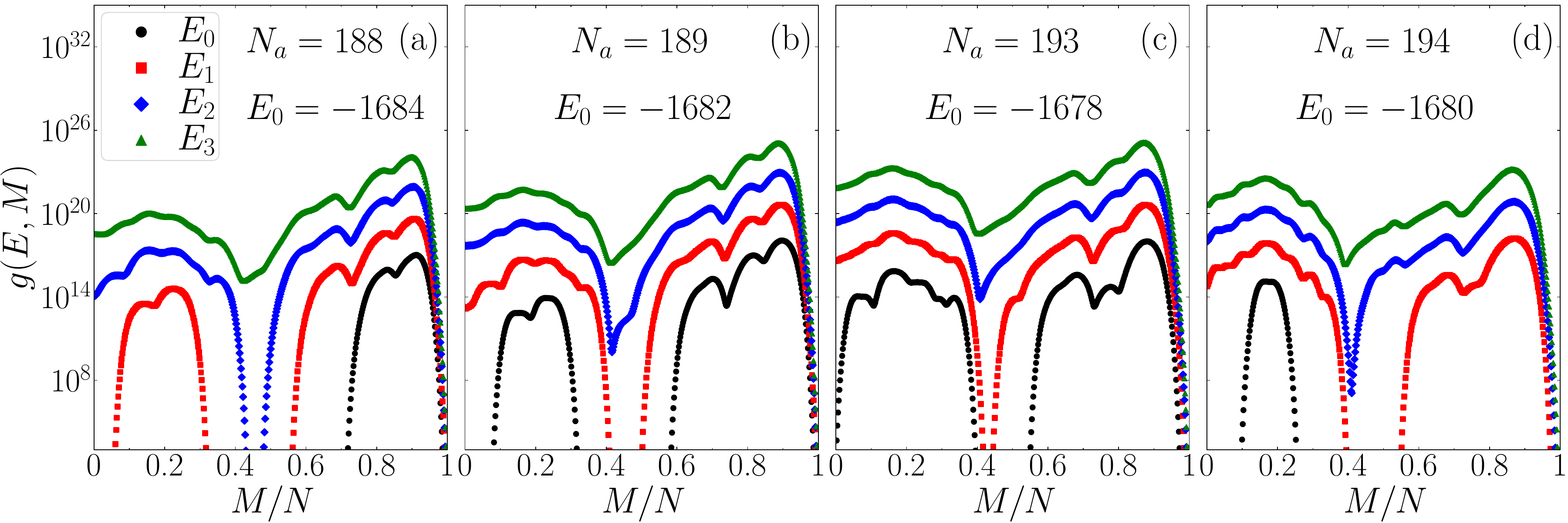} %
    \caption{Density of states \( g(E,M) \) from the ground state \( E_0 \) to the third excited state \( E_3 \), shown for four different values of \( N_a \) in a system of size \( L = 32 \). Each subplot corresponds to a specific \( N_a \). A significant change in the ground-state magnetization distribution is observed between \( N_a = 188 \) and \( N_a = 189 \), where a bimodal structure emerges. Another abrupt transition occurs between \( N_a = 193 \) and \( N_a = 194 \), where the right peak in the ground-state distribution disappears, leaving only the left peak. Both transitions are induced by the addition of a single AFM bond.}
    \label{fig:DOS_at_different_time}
\end{figure*}

Figure~\ref{fig:DOS_at_different_time} shows the DOS \( g(E,M) \) of the ground state and the first, second, and third excited states of a $32\times 32$ system near the transition point $p_{c0}$, following a particular sequence of bond flips. Spin configurations at a given $E$ break up into two well-separated branches: the first branch on the right with only a small fraction of spins in the opposite direction, and the second branch on the left where a substantial fraction of spins are in the opposite direction. For both branches, the spacing between curves at successive $E_n=E_0+4n, n=0,1,\ldots,$ is largely uniform on logarithmic scale, so that peaks at different $E_n$ align at similar magnetization values. 

When the AFM bonds on the lattice are isolated from each other, the ground state is unique, with $M=N$ and an energy $E_g=-2(N-N_a)$. For the case shown in figure~\ref{fig:DOS_at_different_time}(a), one obtains $E_g=-1672$ which differs from the actual $E_0$ by only six bond energies. This indicates that the majority of AFM bonds remain unconnected from each other even as the system approaches the phase boundary.  Flipping either of the two spins connected by an isolated AFM bond increases the system energy by 4. This observation leads to an estimate for the multiplicity of the $n$th excited state 
\begin{equation}
g(E_n,M-2)\simeq 2N_a g(E_{n-1},M),
\label{eq:multiplicity_growth}
\end{equation}
which is in agreement with the observation made above. 

Low energy configurations with $M$ much smaller than $N$ 
cannot be formed by flipping spins connected to isolated AFM bonds. Instead, they arise from flipping an entire cluster of spins, with minimal cost in domain-wall energy. As $N_a$ increases, this process first occurs in the excited states and gradually descends to the ground state. 
Figure~\ref{fig:DOS_at_different_time}(a) and \ref{fig:DOS_at_different_time}(b) 
illustrate how this happens upon the flipping of a single bond: a subset of configurations within the left branch of the first excited state reduce their energy by 2, creating the left branch of the new ground state; on the other hand, ground-state configurations in \ref{fig:DOS_at_different_time}(a) all increase their energy by 2 and making up the right branch of the new ground state. In contrast, the situation depicted in figures \ref{fig:DOS_at_different_time}(c) and \ref{fig:DOS_at_different_time}(d) is slightly different. Upon a single bond flip, the new ground state descends from a subset of configurations in the ground state in \ref{fig:DOS_at_different_time}(c), whereas the rest of previous ground state configurations increase their energy by 2. As a result, the high magnetization branch of the new ground state is eliminated.

More generally, for a given spin configuration, flipping a single bond either increases or decreases the system energy by 2, depending on whether the bond was previously satisfied (i.e., with energy $-1$) or unsatisfied (with energy 1). Consequently, the evolution of $g(E,M)$ can be understood as an exchange of spin configurations between neighboring energy states. Since configurations with smaller $m$ tend to have more nearest-neighbor pairs of opposite spins, their energies are more likely to decrease upon a bond flip compared to those with larger $M$. As a result, for the ground state and low-lying excited states, the magnetization profile $g(E,M)$ at a given $E$ shifts towards smaller $M$ as $N_a$ increases.

\begin{figure}[t]
    \centering
    \includegraphics[width=0.49\linewidth]{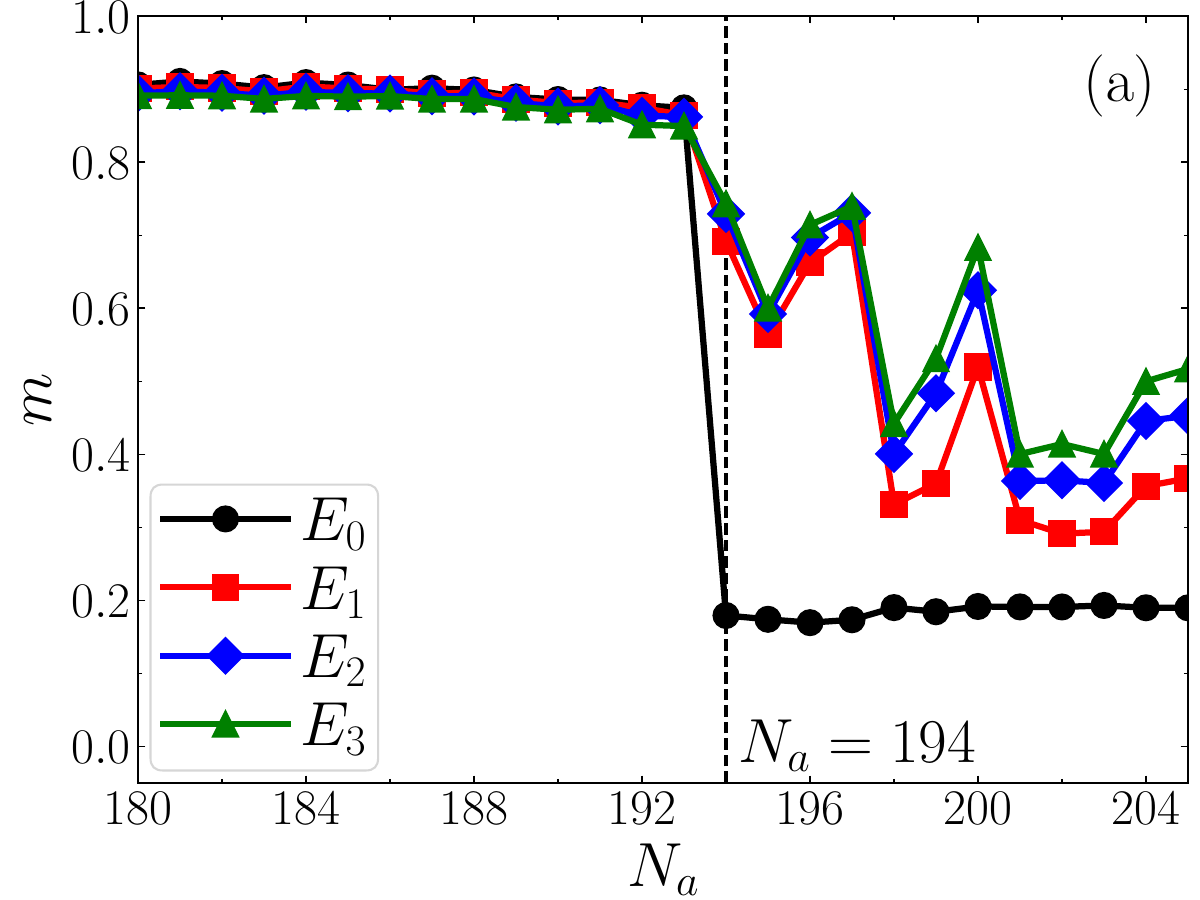} 
    \includegraphics[width=0.49\linewidth]{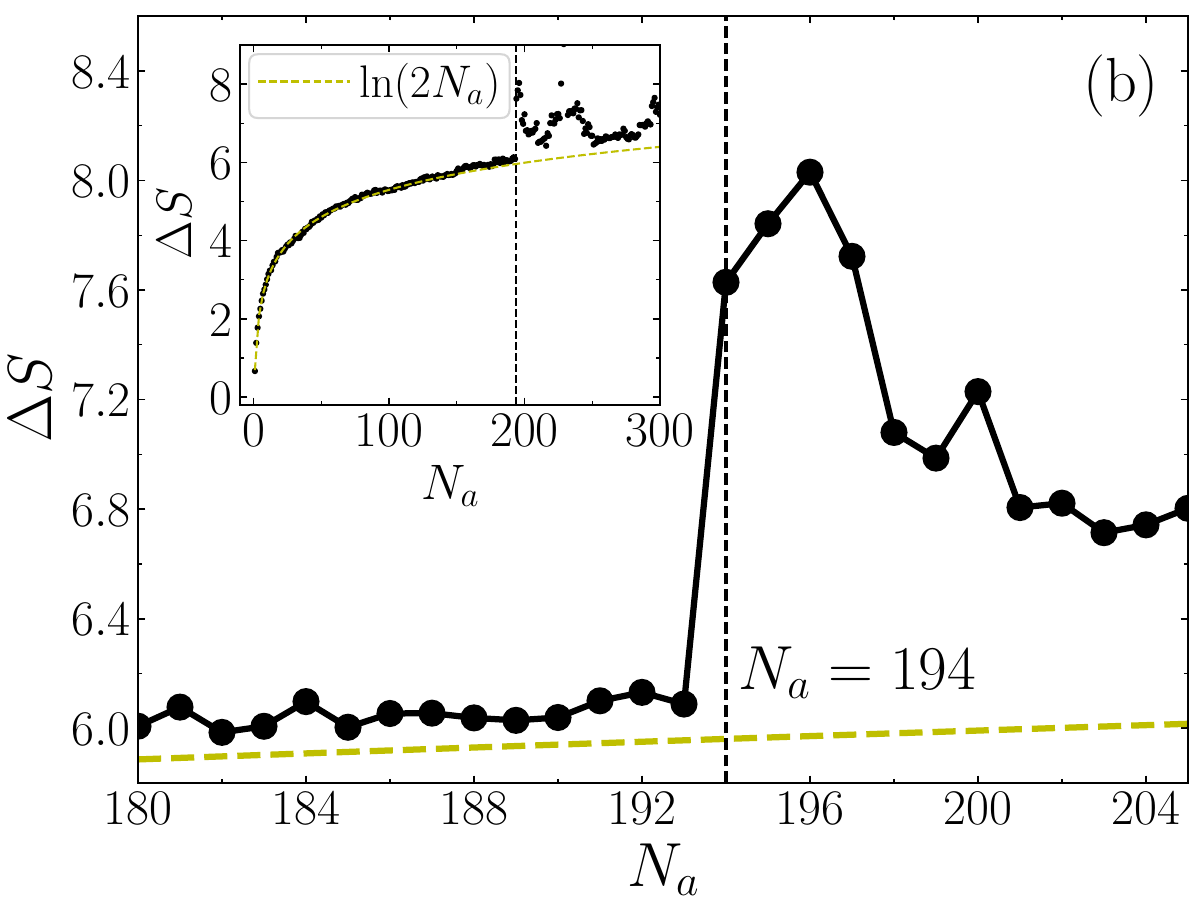} 
    \caption{(a): Average magnetization $m$ as a function of the number of AFM bonds $N_a$ from the ground state ($E_0$) to the third excited state ($E_3$). For this sequence of bond configurations, the ground-state magnetization exhibits a first-order transition. (b): Entropy gap $\Delta S$ between the first excited state and the ground state versus $N_a$. The dashed line, $\ln(2N_a)$, represents the expected contribution from completely independent (single spin) excitations. A sudden increase in $\Delta S$ at $N_a = 194$ suggests a change in the excitation mechanism at this point.}
    \label{fig:M vs t}
\end{figure}

Figure~\ref{fig:M vs t}(a) shows the evolution of average magnetization for low-lying states as $N_a$ increases. Up till \( N_a < 194 \), these states remain FM with \( m(E_n) > 0.8 \). At \( N_a = 194 \), an abrupt drop in ground-state magnetization (\( \Delta m \approx 0.7 \)) is observed, due to loss of the high magnetization branch (figure~\ref{fig:DOS_at_different_time}(d)). In contrast, the magnetization of the excited states show only a moderate decrease (\( \Delta m \approx 0.2 \)), without losing their high magnetization branch at this point. 

Figure~\ref{fig:M vs t}(b) shows the entropy gap $\Delta S=S_1-S_0=\ln[g(E_1)/g(E_0)]$
between the first excited and ground states against $N_a$ in the neighborhood of the transition. Based on equation~\eqref{eq:multiplicity_growth}, we expect $\Delta S \simeq \ln (2N_a)$ at sufficiently small $N_a$, which is borne out by data in the inset up till very close to the transition.
This confirms that dominant excitations above the FM ground state are single spin flips. At $N_a=194$, however, $\Delta S$ increases significantly, indicating that more complex excitations emerge. The somewhat chaotic behavior of $\Delta S$ to the right is attributed to large fluctuations in $g(E_0)$ around the transition.

\begin{figure}[t]
    \centering
    \includegraphics[width=0.8\linewidth]{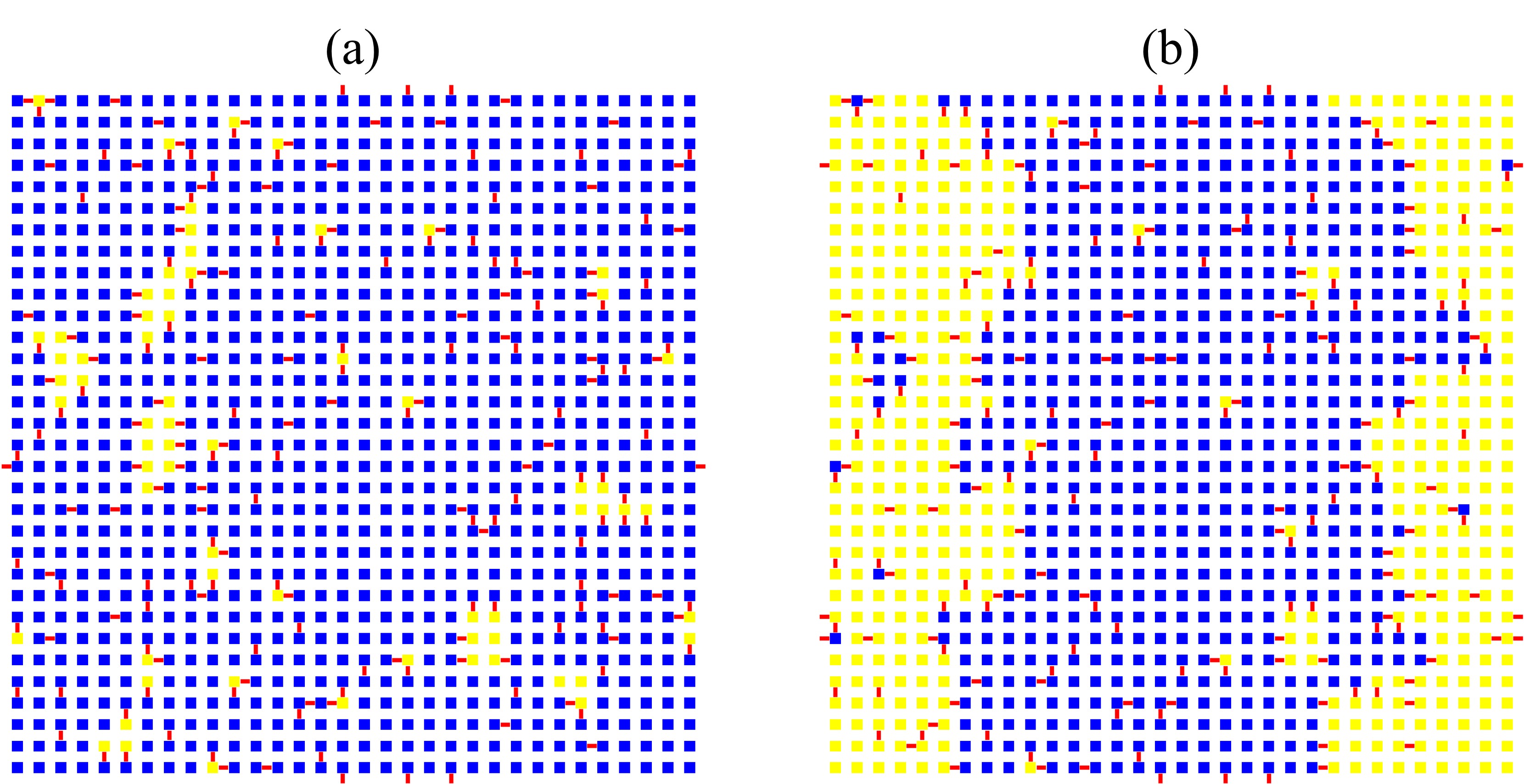} 
    \caption{Spin and bond configurations in the ground state for $N_a = 193$ (a) and $N_a = 194$ (b). Blue and yellow cubes denote spins $S_i = 1$ and $S_i = -1$, respectively, while red bonds indicate frustrated bonds with $J_{ij}S_iS_j = -1$. For $N_a = 193$, both (a) and (b) represent degenerate ground states, whereas for $N_a = 194$, only configuration (b) remains as the ground state. The primary difference between the two configurations is the presence of a large spin cluster}
    \label{fig:Conf_gs}
\end{figure}

To investigate the microscopic origin of this abrupt change, we compare ground-state spin configurations at the largest $g(E_0,M)$ before and after the transition. Figure~\ref{fig:Conf_gs}(a) shows the most probable configuration at $N_a=193$: a nearly uniform ferromagnet configuration with only a few isolated spins in the opposite direction. 
By contrast, panel (b) displays the ground state at $N_a=194$, which is no longer a single domain but instead splits into two oppositely magnetized regions of comparable size. The sudden emergence of multi-domain configurations underlies the observed collapse of the net magnetization and the concurrent jump in entropy. We have confirmed that the observations described above hold generally for different sequences of bond flips, although the precise value of $N_a$ to destroy the FM ground state differs from sample to sample.

\subsection{Finite-size scaling analysis of ground-state magnetization}
\label{sec:M_in_gs}

\begin{figure*}[t]
    \centering
    \includegraphics[width=\linewidth]{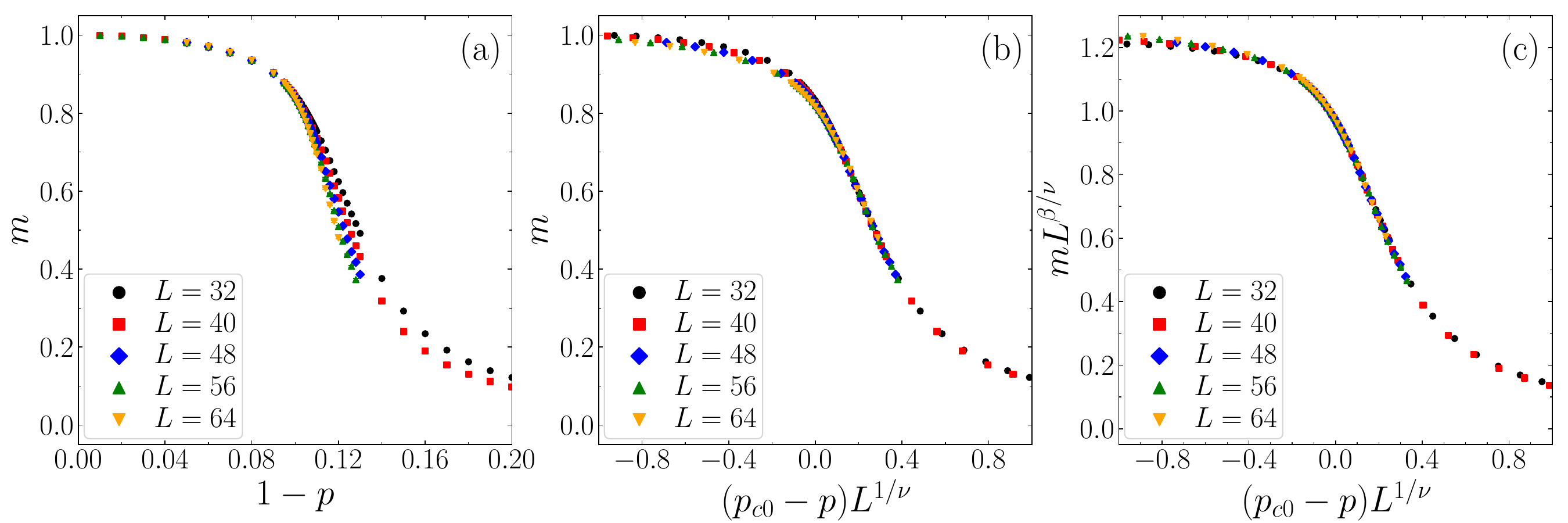} %
    \caption{Average ground-state magnetization $m$ and its FSS under different parameters. (a) Ground-state magnetization $m$ as a function of $1-p$. (b) and (c): FSS analysis of $m$ using $(p_{c0}, \beta) = (0.898, 0)$ in (b) and $(p_{c0}, \beta) = (0.8945, 0.083)$ in (c).}
    \label{fig:M_vs_p}
\end{figure*}

In section~\ref{sec:fss_results}, we perform a standard FSS analysis of the zero-temperature Binder cumulant and magnetization, yielding critical exponents $\nu = 1.50(8)$ and $\beta = 0.083(5)$. Based on these values, the data collapse of the ground-state magnetization, shown in figure~\ref{fig:M_vs_p}(c), exhibits excellent quality. However, the resulting scaling behavior does not align with our anticipated physical picture.

We propose our FSS analysis based on our physical picture. Before the transition, the system behaves exactly as a ferromagnet, a sudden decay of the ground-state magnetization is expected only after the transition. Thus we consider two scaling regions for magnetization near the critical point:
\begin{equation}
    m(p, L) =
    \begin{cases} 
    m(p), & p > p_{c0} \\
    \tilde{m}((p - p_{c0}) L^{1/\nu}), & p < p_{c0}
    \end{cases}
    \label{eq:m_scaling_new}
\end{equation}
We perform a data collapse of the segmented scaling function using \( p_{c0} = 0.898 \) and \( \nu = 1.5 \), as shown in figures~\ref{fig:M_vs_p}(a) and (b). The quality of the collapse is visually excellent and appears comparable to that obtained from the standard FSS analysis (figure~\ref{fig:M_vs_p}(c)).

To further examine the scaling behavior, we fix \( \nu = 1.5 \) and fit the ground-state magnetization data in the range \( 0.102 < 1 - p < 0.112 \) using the following form:
\begin{equation}
     m = \sum_{n=0}^{n_{\max}} a_n (p - p_{c0})^n L^{n/\nu} 
    \label{eq:M_fitting_two_region}
\end{equation}
We gradually increase the minimum system size \( L_{\text{min}} \) to assess the quality of the fit. The corresponding results are summarized in Table~\ref{tab:M_scaling_zero_beta}. As \( L_{\text{min}} \) increases, \(\chi^2/\text{DOF}\) approaches 1, indicating that \(\beta = 0\) is reasonable in the FSS analysis for larger system sizes. Based on the results in Table~\ref{tab:M_scaling_zero_beta}, we conservatively estimate the zero-temperature critical point to be \( p_{c0} = 0.898(2) \).

\begin{table}[t]
\centering
\caption{Fitting results for the FSS of the magnetization \( m \) at different $L_{\min}$, using the equation~\eqref{eq:M_fitting_two_region} by fixing $\beta=0$ and $\nu=1.5$.}
\label{tab:M_scaling_zero_beta}
\begin{tabular}{ccc}
$L_\text{min}$ & $\chi^2/\text{DOF}$ & $p_{c0}$  \\  
 \hline
$24$  &  $3345.9 / 69$  &  $0.9008(4)$  \\
$32$  &  $974.2 / 53$  &  $0.9001(4)$  \\
$40$  &  $217.6 / 37$  &  $0.8991(7)$  \\
$48$  &  $99.6 / 31$  &  $0.8988(6)$  \\
$56$  &  $16.5 / 13$  &  $0.8982(12)$  \\
\end{tabular}
\end{table}

\begin{figure*}[t]
    \centering
    \includegraphics[width=\linewidth]{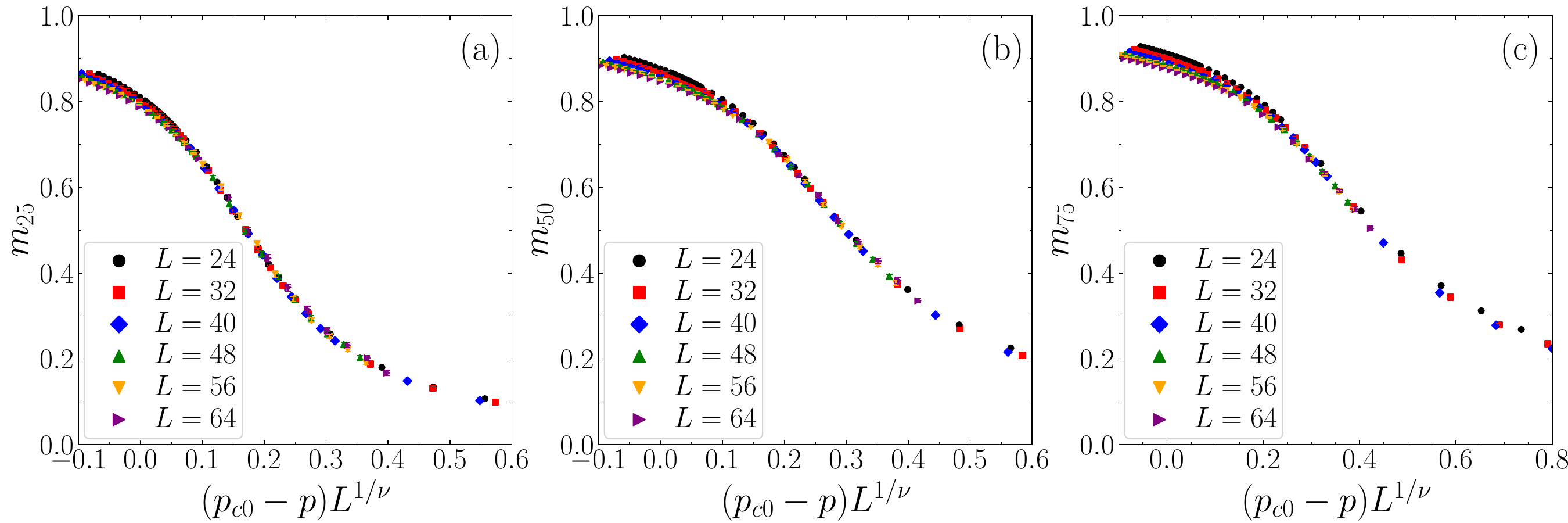} %
    \caption{FSS analysis of ground-state magnetization quartiles, performed with fixed critical exponents $\beta = 0$ and $\nu = 1.5$. Panels (a)–(c) display scaling collapses for the lower quartile ($m_{25}$), median ($m_{50}$), and upper quartile ($m_{75}$), respectively. (a) $m_{25}$: Best-fit critical point $p_{c0} = 0.8969(6)$, minimum system size $L_{\min} = 24$, and goodness-of-fit $\chi^2/\mathrm{DOF} = 81.1/60$.(b) $m_{50}$: $p_{c0} = 0.8980(7)$, $L_{\min} = 32$, $\chi^2/\mathrm{DOF} = 47.9/36$. (c) $m_{75}$: $p_{c0} = 0.898(2)$, $L_{\min} = 40$, $\chi^2/\mathrm{DOF} = 25.1/15$.}
    \label{fig:M_quar}
\end{figure*}


In addition to the average ground-state magnetization, we analyze the quartiles of the magnetization as a characterization of its distribution. Near the zero-temperature critical point, the quartile magnetization \( m_q \) exhibits scaling behavior analogous to that of the average magnetization. 

We define the lower quartile \( m_{25} \), the median quartile \( m_{50} \), and the upper quartile \( m_{75} \) of the magnetization based on the cumulative distribution function of the magnetization distribution \( P(m, p, T) \), which is normalized such that $\sum_{m} P(m, p, T) = 1$. The cumulative distribution function is given by $Q(m, p, T) = \sum_{|m'| \le |m|} [P(m', p, T)]$.
Then, the quartiles are defined as the magnetization values satisfying:
\begin{align}
    Q(m_{25}, p, T) &= 0.25, \label{eq:M_25} \\
    Q(m_{50}, p, T) &= 0.50, \label{eq:M_50} \\
    Q(m_{75}, p, T) &= 0.75. \label{eq:M_75}
\end{align}
The interquartile range (IQR) of the magnetization is defined as
\begin{equation}
    m_{IQR} = m_{75} - m_{25}.
    \label{eq:M_IQR}
\end{equation}

\begin{figure}[t]
    \centering
    \includegraphics[width=0.49\linewidth]{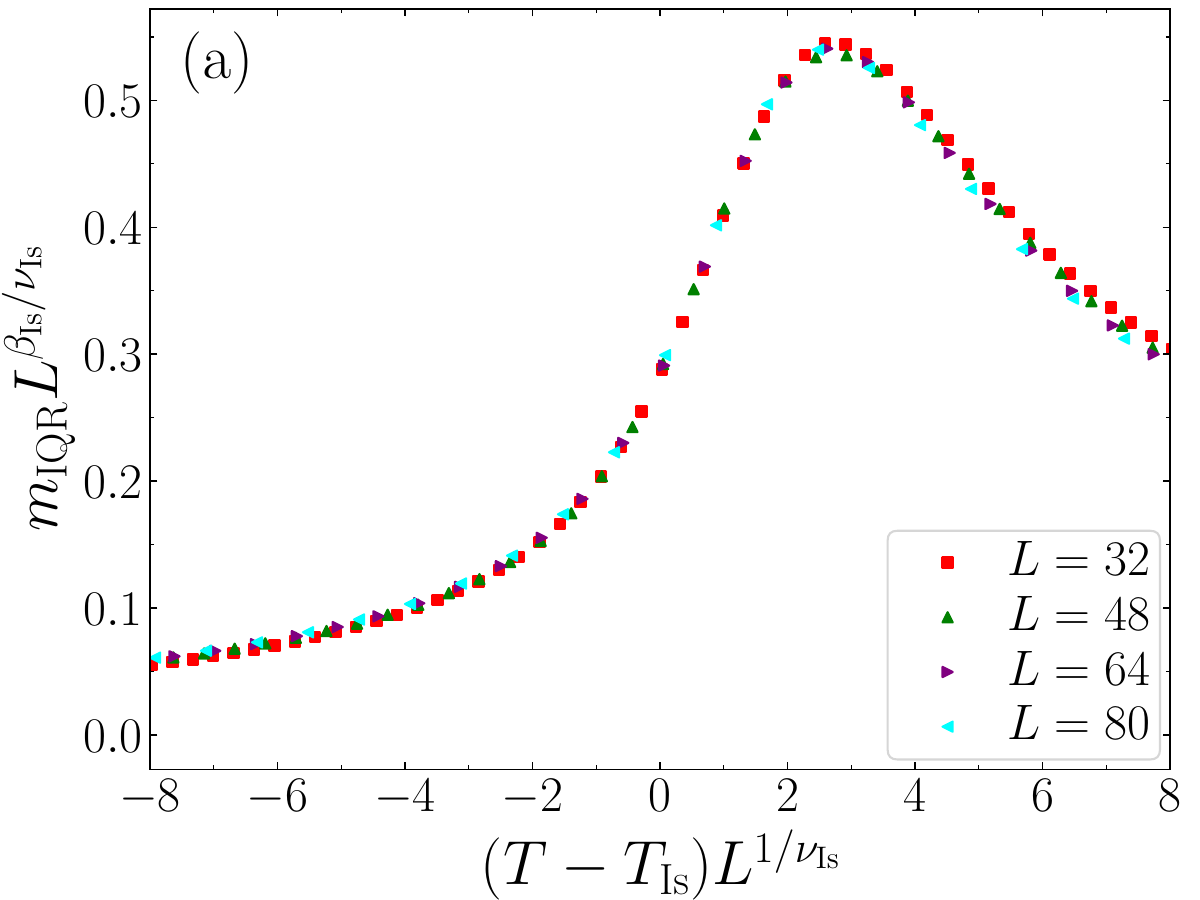} %
    \includegraphics[width=0.49\linewidth]{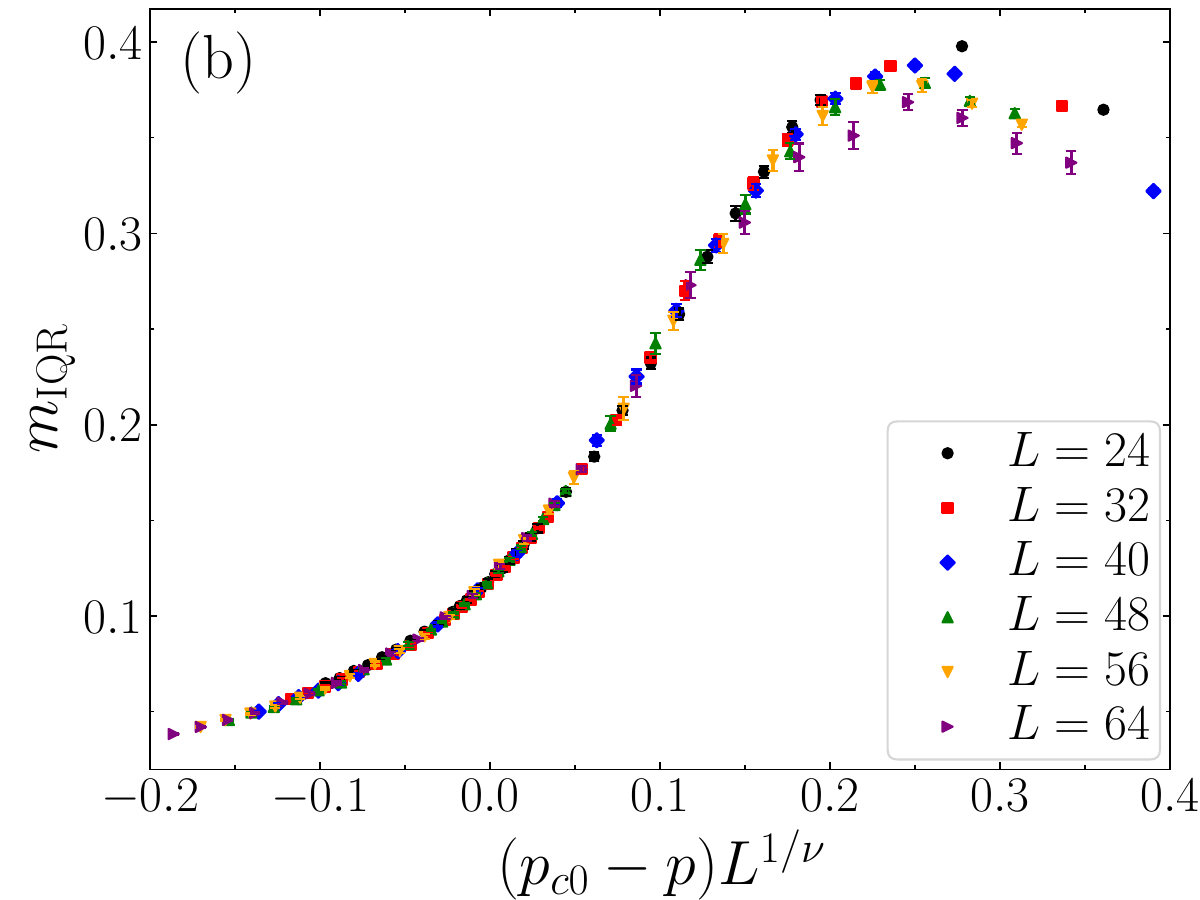} %
    \caption{FSS analysis of the IQR of magnetization at Ising critical point and zero-temperature critical point. (a): FSS at Ising critical point using $T_{\text{Is}} = 2.269$, $\beta_{\text{Is}} = 1/8$, and $\nu_{\text{Is}} = 1$. (b): FSS at zero-temperature critical point, with best-fit $p_{c0} = 0.893(5)$ and goodness-of-fit $\chi^2/\mathrm{DOF} = 88.5/77$, using fixed exponents $\beta = 0$, $\nu = 1.5$, and minimum system size $L_{\min} = 24$. }
    \label{fig:M_IQR}
\end{figure}

The same scaling analysis procedure as described in the analysis of ground-state magnetization in this section is employed, with the critical exponents fixed at \(\beta = 0\) and \(\nu = 1.5\). To properly account for the FSS behavior on the SG side of the transition, only data satisfying \( m_q < 0.7 \) are considered. Data with \( m_q > 0.7 \) are predominantly influenced by contributions from the FM order, especially in the case of \( m_{75} \). As shown in figure~\ref{fig:M_quar}, the data exhibit a good collapse in the SG phase for the lower, median, and upper quartiles, denoted \( m_{25} \), \( m_{50} \), and \( m_{75} \), respectively. The corresponding estimates of the critical point are \( p_{c0}= 0.8969(6) \), \( p_{c0} = 0.8980(7) \), and \( p_{c0} = 0.898(2) \), respectively.

The quasi-first-order nature of the zero-temperature transition suggests the existence of a bimodal structure in the ground-state magnetization distribution in the thermodynamic limit, indicative of phase coexistence. This feature is also observed in figure~\ref{fig:DOS_at_different_time}. 
The upper and lower quartiles, \( m_{75} \) and \( m_{25} \), are identified as the right and left peaks of the distribution, respectively. The IQR of the magnetization, $m_{\text{IQR}}$, serves as a measure of the separation between the two peaks.
A finite value of \( m_{\text{IQR}} \) in the thermodynamic limit constitutes evidence for a bimodal magnetization distribution.

We compare the behavior of \( m_{\text{IQR}} \) at two critical points of the 2D $\pm J$ RBIM. At the pure Ising critical point, which corresponds to a well-established second-order phase transition, the IQR of magnetization exhibit excellent FSS collapse at the critical point when rescaled using the exact critical exponents, as shown in figure~\ref{fig:M_IQR}(a). The data of \( m_{\text{IQR}} \) scales as \( L^{-1/8} \) and vanishes in the thermodynamic limit, indicating the absence of a bimodal distribution at the critical point.
In contrast, at the zero-temperature critical point, the data collapse well with \(\beta = 0\) and \(\nu = 1.5\), and a critical point \(p_{c0} = 0.893(5)\), as shown in figure~\ref{fig:M_IQR}(b). The critical exponent \(\beta = 0\) implies that \( m_{\text{IQR}} \) does not vanish with increasing system size, but instead converges to a finite value. This strongly suggests the presence of a bimodal structure in the ground-state magnetization distribution in the thermodynamic limit, providing direct evidence that the transition is quasi-first-order in nature.

\section{Discussion}
\label{sec:discussion}
In this work, we investigated the 2D \( \pm J \) RBIM using the entropic sampling method with an efficient iterative algorithm that we implement. Our algorithm has been successfully applied to the 2D $\pm J$ RBIM, reaching system sizes up to \( L = 64 \),  allowing for detailed FSS analysis of the Binder cumulant and magnetization.

From these analyses, we determined the reentrant phase boundary below the MCP and extracted the associated critical exponents \( \nu \) and \( \beta \). Detailed FSS results are summarized in Tables~\ref{tab:B_scaling} and~\ref{tab:M_scaling}. The exponent \( \nu \) varies between 1.50(8) and 1.56(6) below the MCP, suggesting that the zero-temperature critical point and the MCP share the same correlation length exponent. The magnetization exponent \( \beta \) decreases from 0.134(5) at \( T = 0.95 \) to 0.083(5) at zero temperature. The critical point \( p_c \) shifts from $0.89075(8)$ at \( T = 0.95 \) to $0.89456(13)$ at \( T = 0 \). 

By analyzing magnetization in the ground state and low-energy excited states, we observed clear signatures of a first-order phase transition in finite-size systems, marked by a sudden drop in magnetization upon introducing a single AFM bond. Further analysis of the DOS, entropy gap and spin configurations of this sample reveals that this drop corresponds to the disappearance of the ground-state high-magnetization branch and the emergence of multi-domain structures.  Based on these results, we argue that the FM-SG transition at zero temperature is a quasi–first-order phase transition.
By considering two distinct scaling regions, we find that setting \(\beta = 0\) in the FSS analysis of both the average magnetization and its quartiles is justified. This result supports the interpretation that the zero-temperature transition is quasi–first-order in nature, in the sense that the ground-state magnetization experiences a sudden decay only after the transition in the thermodynamic limit, where a critical exponent $\nu=1.5$ is observed on the SG side. 
The presence of a bimodal magnetization structure near the zero-temperature critical point was confirmed via FSS of the IQR of magnetization. 
Based on the FSS analysis of the ground-state magnetization \(m\) and its quartiles \(m_{25}\), \(m_{50}\), \(m_{75}\), we conservatively estimate the zero-temperature critical point to be \(p_c = 0.898(2)\).


Our findings at the zero-temperature critical behavior deviate slightly from previous studies ~\cite{amoruso2004domain}, where exact matching algorithms were used to simulate systems up to $L=700$. Although their method accurately determines the ground state, magnetization was computed using zero-temperature single-spin-flip dynamics, which may not fully capture the bimodal nature of the magnetization distribution near the critical point. This limitation could lead to inaccuracies in magnetization-related observables.

\section*{Acknowledgments}
We thank Leticia Cugliandolo for many helpful discussions during the course of the project. 
We acknowledge the High-Performance Computing Center at Westlake University for supporting the simulations.
This work is supported by NKRDPC-2022YFA1402802, NSFC-92165204, the Research Grants Council of the HKSAR under grants 12304020 and 12301723, Guangdong Provincial Key Laboratory of Magnetoelectric Physics and Devices (No. 2022B1212010008), Guangdong Fundamental Research Center for Magnetoelectric Physics, and Guangdong Provincial Quantum Science Strategic Initiative (GDZX2401010).

\appendix

\section{Simulation details}
\label{sec:simulation_detail}

\begin{table}[t]
\centering
\caption{\( N_s \) represents the number of disorder realizations. A large number of disorder averages were performed for $L = 24,~32,~48$, and 64 within \( 0.89 \leq p \leq 0.896 \) to achieve high-precision FSS analysis in section~\ref{sec:fss_results}.}
\label{tab:Ns}
\renewcommand{\arraystretch}{1.2} 
\setlength{\tabcolsep}{8pt} 
\begin{tabular}{cc|cc} 
\toprule
\multicolumn{2}{c|}{$0.89 \leq p \leq 0.896$} & \multicolumn{2}{c}{The rest of $p$} \\
\cmidrule(lr){1-2} \cmidrule(lr){3-4}
$L$ & $N_s/10^3$ & $L$ & $N_s/10^3$ \\
\midrule
$24$  & $1000$ & $24$  & $100$ \\
$32$  & $1000$ & $32$  & $100$ \\
$40$  & $30$   & $40$  & $30$  \\
$48$  & $300$  & $48$  & $30$  \\
$56$  & $5$    & $56$  & $5$   \\
$64$  & $50$   & $64$  & $5$   \\
\bottomrule
\end{tabular}
\end{table}

We focus on the phase behavior near and below the MCP of the 2D $\pm J$ RBIM. 
For regions below the MCP and system sizes \( L \geq 16 \), we observe that only a subset of the energy levels contributes significantly. Truncating the energy range to \([E_0, E_0 + N/4]\) yields results identical to those obtained with the full energy range for \( T \leq 1.0 \), according to our tests. Since the exact ground-state energy is unknown at the beginning of the simulation, we use a local energy minimum \( E'_0 \) and restrict the sampling to energies below \( E'_0 + N/4 \). This approach significantly reduces computational cost without affecting the final results. 


\begin{figure}[t]
    \centering
    \includegraphics[width=0.7\linewidth]{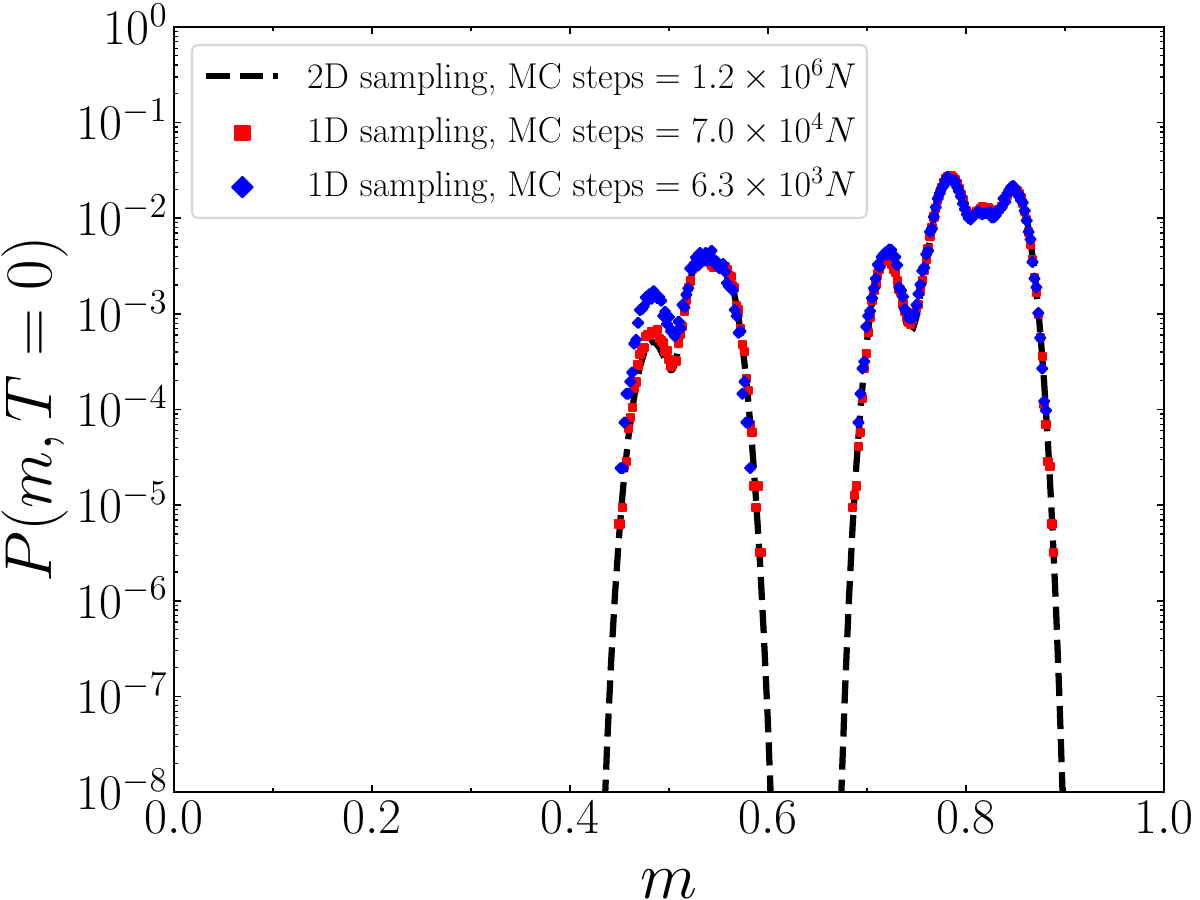}
    \caption{Comparison of 1D and 2D sampling in estimating the ground-state magnetization distribution for a representative sample with \( N = 32 \times 32 \) and \( p = 0.88 \). The 1D sampling converges to \( h_{\min}/h_{\max} > 0.7 \) with only \( 6.3 \times 10^3 N \) MC steps, while achieving \( h_{\min}/h_{\max} > 0.95 \) requires approximately \( 7.0 \times 10^4 N \) steps. The 2D sampling yields highly accurate distributions at a much higher computational cost.}
    \label{fig:P_M_compare}
\end{figure}

In all simulations, convergence is determined by the histogram flatness criterion \( h_{\min}/h_{\max} \geq 0.7 \), where \( h_{\min} \) and \( h_{\max} \) denote the minimum and maximum values of the histogram, respectively. 
Once the DOS has converged, we perform a production run with MC steps (one MC step corresponds $N$ spin flip attempts) equal to twice the number used in the final iteration that satisfied the histogram flatness criterion, in order to accumulate high-resolution histograms of magnetization at each energy level.

The number of disorder realizations for different system sizes is listed in Table~\ref{tab:Ns}.
To measure physical quantities over a large number of disorder realizations, we perform energy-space sampling (one-dimensional (1D) sampling), obtaining the DOS \( g(E) \) using our iterative scheme. 
For the analysis of individual samples (section~\ref{sec:reentrant_sample}), we employ 2D sampling, determining the DOS \( g(E, M) \) with Berg's iterative scheme.

Through our tests, we found this to be a promising approach that balances computational efficiency and accuracy. Figure~\ref{fig:P_M_compare} presents the test results for a representative sample with system size \( N = 32 \times 32 \) and  \( p = 0.88 \). The 2D sampling method yields highly accurate ground-state magnetization distributions, but at a prohibitive computational cost. In contrast, 1D sampling with a stringent convergence criterion of \( h_{\min}/h_{\max} > 0.95 \) achieves comparable accuracy with the 2D sampling method in the regime where \( P(m) > 10^{-5} \). However, in practice, only the region with \( P(m) > 10^{-3} \) contributes significantly to physical observables. We find that using a relaxed convergence criterion of \( h_{\min}/h_{\max} > 0.7 \) already provides a very good approximation, with negligible impact on both the average magnetization and its quartiles.



\bibliographystyle{unsrt}
\bibliography{reference}

\end{document}